\documentclass[a4paper]{jpconf}
\usepackage{color}
\usepackage{graphicx}
\begin{document}
\title{Rapidity dependence of azimuthal correlations for p+p and d+Au at  $\sqrt{s}=200$GeV}

\author{Xuan Li$^{[1,2]}$ for the STAR Collaboration}

\address{1) School of Physics, Shandong University, Jinan, 250100, China}
\address{2) Brookhaven National Lab, Bldg. 510A, Physics Department, upton, NY, 11973, USA}

\ead{xuanli@rcf.rhic.bnl.gov}

\begin{abstract}
Forward di-jet production in d+Au collisions at RHIC provides sensitivity to Bjorken-x between 0.001 and 0.02 for the nuclear gluon density . The STAR experiment at RHIC has continuous azimuthal coverage with electromagnetic calorimeters for pseudo-rapidity from -1 to 4.  We have measured forward+mid-rapidity $\pi^{0}-\pi^{0}$ and $\pi^{0}-h$ azimuthal correlations and forward+forward rapidity $\pi^{0}-\pi^{0}$ azimuthal correlations. Both analyses use the Forward Meson Spectrometer (FMS) triggered data in RHIC 2008 run p+p and d+Au collisions at $\sqrt{s_{NN}}$=200 GeV. The end-cap electromagnetic calorimeter (EEMC) at STAR covers pseudo-rapidity from 1.07 to 2. This provides us sensitivity to the nuclear gluon density at intermediate x values. Preliminary results on the $\pi^{0}-\pi^{0}$ and $\pi^{0}$ + jet-like azimuthal correlations with both the FMS and the EEMC will be discussed.
\end{abstract}

\section{Introduction}
Deep inelastic scattering experiments show that the nucleon gluon density increases rapidly as Bjorken-x ($x_{BJ}$) decreases\cite{dis}. The increase at low $x_{BJ}$ is due to partons splitting into more partons. As the gluon density increases, two partons recombining into one becomes increasingly likely. When parton recombination compensates parton splitting, the number of partons will stop increasing and parton saturation is realized. At a given $x_{BJ}$, the nuclear (mass number A) gluon density is approximately $A^{1/3}$  times the nucleon gluon density\cite{refimpact}. Consequently, saturation in the nuclear gluon density is expected to be realized at higher $x_{BJ}$ region than for the nucleon gluon density at fixed $Q^{2}$. 

Forward inclusive $\pi^{0}$ production probes asymmetric partonic scattering and primarily is from a large $x_{F}$ quark scattering from a low $x_{BJ}$ gluon. The inclusive production measured in p+p collisions at $\sqrt{s}=200$GeV is consistent with NLO perturbative Quantum Chromodynamics (pQCD) descriptions. Inclusive production measured in d+Au collisions at $\sqrt{s}=200$GeV are better described by Color Glass Condensate (CGC) models\cite{refinclusive, refthe1}. The lower x region can be selectively probed through forward di-jet correlations\cite{refthe2}. The shape and magnitude of back-to-back azimuthal correlation is related to the parton density distribution\cite{for_for,refmono}. 

With RHIC 2008 run p+p and d+Au collisions, we have measured both forward+mid-rapidity and forward+forward rapidity azimuthal correlations to probe low $x_{BJ}$ nuclear parton distributions. From the forward+mid-rapidity $\pi^{0}$-$\pi^{0}$ and $\pi^{0}-h$ measurements, the back-to-back azimuthal correlation peaks look similar in p+p and d+Au data. Lack of significant broadening from p+p to d+Au indicates the forward+mid rapidity correlations are not near the saturation region\cite{for_cen}. Broadening from p+p to d+Au has been observed in the forward+forward rapidity back-to-back azimuthal correlations, especially in the central dAu data which show good agreement with the CGC predictions\cite{for_for}. The observed suppression of the away-side peak in central d+Au forward di-pion events is consistent with gluon saturation expectations. 

With $\pi^{0}$ measured in the forward region, the transition to saturation can be probed through varying the pseudo-rapidity of the associated particle (or jet-like clusters). Azimuthal angle correlations between particles detected at forward rapidity associated with particles at near-forward rapidity probe the intermediate $x_{BJ}$ region to look for the sharpness of the transition from low to high partonic densities\cite{phase}. 
 
\section{Experiment setup}
\begin{figure}[t]
\centerline{\includegraphics[width=85mm]{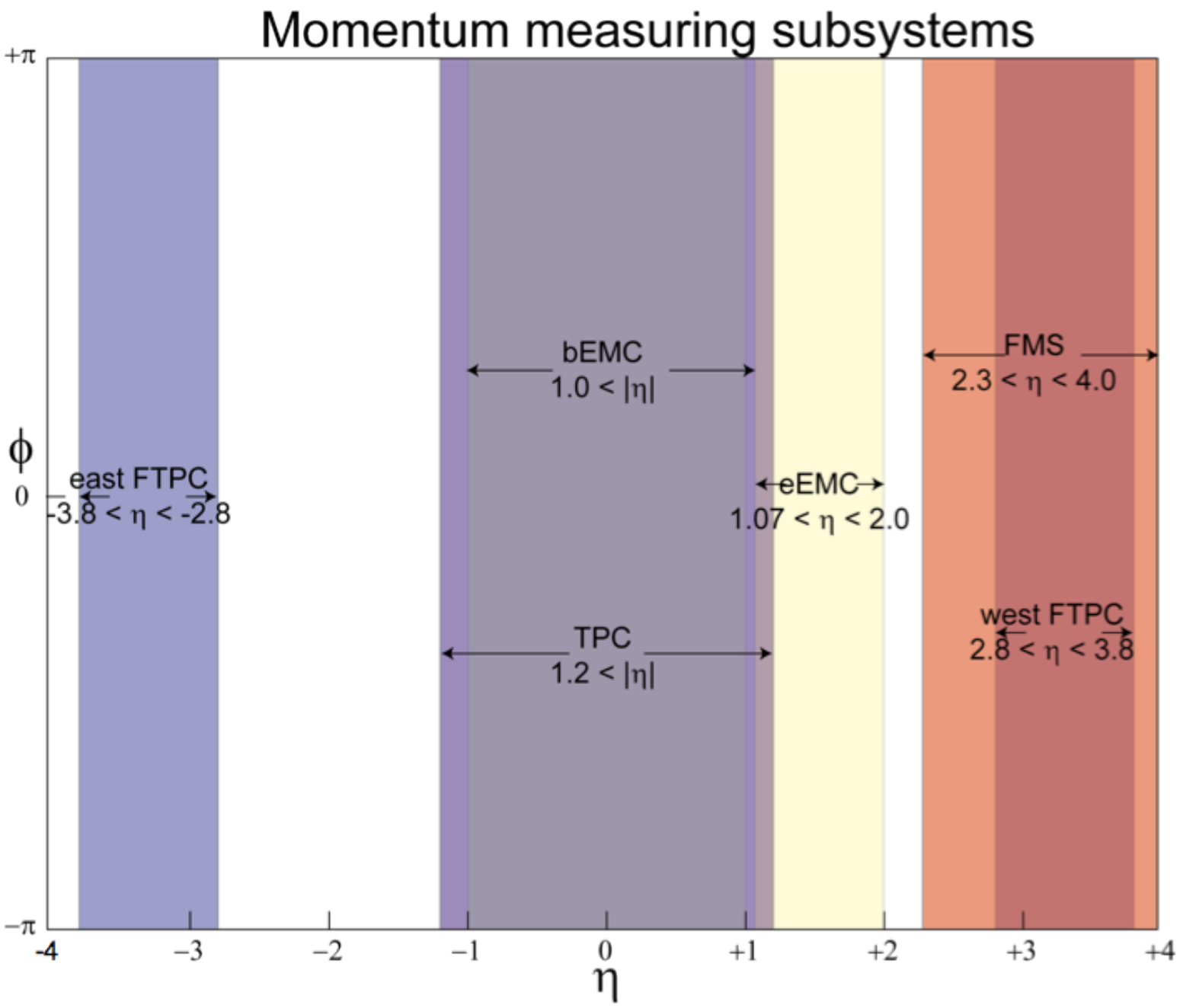}}
\caption{{\it STAR electromagnetic calorimetry $\eta$ $\phi$ coverage.}}
\label{full}
\end{figure}

The Solenoid Tracker at RHIC (STAR) has nearly continuous full azimuth acceptance of electromagnetic calorimeters in $-1.0 < \eta < 4.0 $ (See Figure \ref{full}) . The Barrel ElectroMagnetic Calorimeter (BEMC: $-1.0 < \eta < 1.0$), the End-cap ElectroMagnetic Calorimeter (EEMC: $1.07 < \eta < 2.0 $) and the Forward Meson Spectrometer (FMS: $2.3 < \eta < 4.0$) comprise the STAR electromagnetic calorimeters. The electromagnetic calorimeters together with the Time Projection Chamber (TPC: $-1.0 < \eta < 1.0$) provide a broad $\Delta\eta\times\Delta\varphi$ coverage for correlation measurements.

In the 2008 run, RHIC provided high luminosity p+p and d+Au collisions. The p+p runs provide reference for di-jet correlations in d+Au runs. The FMS, facing the deuteron beam for the d+Au runs, can probe the gluon density in the nuclear down to $x_{BJ} \approx 10^{-4}$. A high-tower trigger was used for the FMS data taken. The forward+near-forward rapidity azimuthal correlations use neutral pions measured in the FMS and neutral pions or jet-like candidates detected in the EEMC.

 \section{Azimuthal correlation analysis}
 Non-linear effects on particles produced when a dilute system (e.g. the deuteron) interacts with dense nuclear targets have been predicted by many models.  One model which is based on pQCD calculations predicts the partons interact coherently with multiple nucleons changing from a $2 \rightarrow 2$ partonic elastic scattering process to $2 \rightarrow many$ process\cite{mulqiu}. With higher gluon density in d+Au interactions than in p+p, CGC models predict broadening of back-to-back azimuthal correlations and their suppression (i.e. mono-jets) due to $2 \rightarrow many$ partonic scattering\cite{refmono}.  The CGC signatures in the low $x_{BJ}$ range get enhanced in the large pseudo-rapidity region. With a forward pseudo-rapidity $\pi^{0}$ triggered in the FMS, we look at the coincident   $\pi^{0}$ in the EEMC. To minimize particle selection bias effect, the FMS $\pi^{0}$ and EEMC jet-like candidate correlations are also studied. 

\begin{figure}[t]
\centerline{\includegraphics[width=145mm]{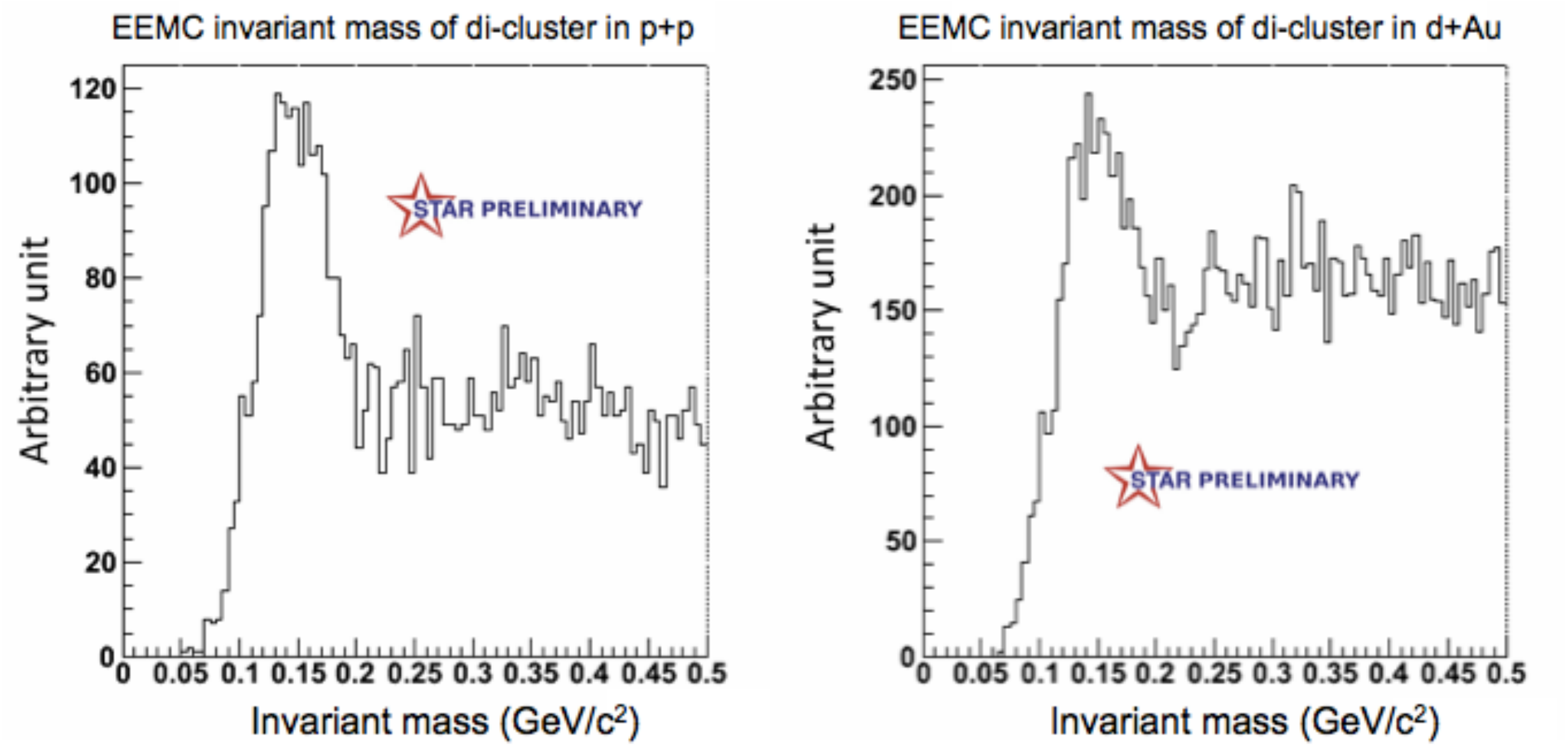}}
\caption{ {\it Invariant masses of EEMC di-clusters ($1.5GeV/c<p_{t}^{EEMC}<p_{t}^{FMS}$) associated with forward $\pi^{0}$ ($p_{t}^{FMS}>2.5GeV/c$) with further event selections mentioned in the text for p+p (left) and d+Au (right) data. }}
\label{pi_mass}
\end{figure}

\begin{figure}[t]
\centerline{\includegraphics[width=145mm]{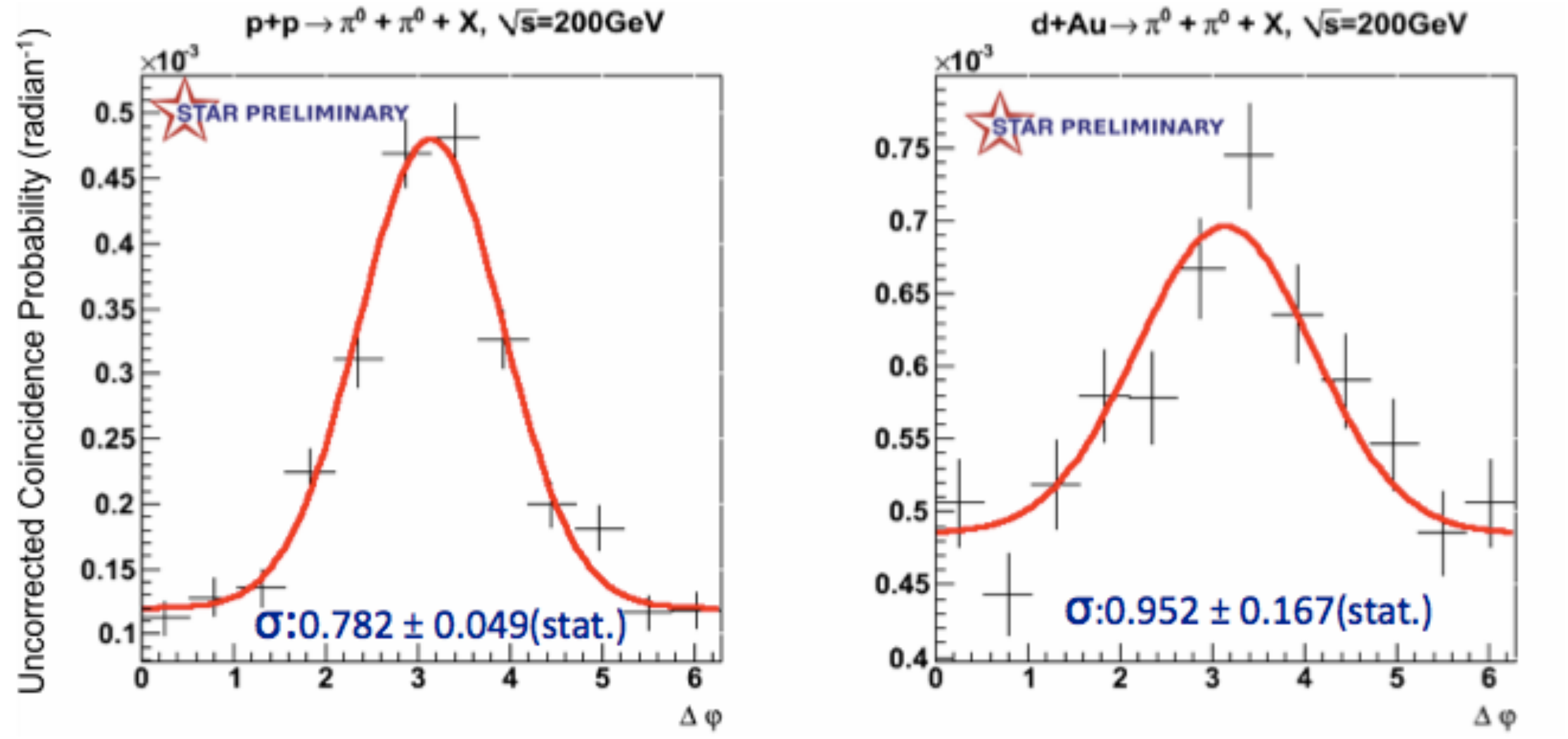}}
\caption{ {\it Uncorrected coincidence probability versus azimuthal angle difference between a forward $\pi^{0}$ ($p_{t}^{FMS}>2.5GeV/c$) and a near-forward rapidity $\pi^{0}$ ($1.5GeV/c<p_{t}^{EEMC}<p_{t}^{FMS}$). The red lines stand for the fitting functions which are a constant plus a Gaussian function. The left plot shows the correlations in p+p data and d+Au results are shown on the right plot. }}
\label{eemc_pi}
\end{figure}

%% pi0 with di-cluster
An electromagnetic calorimeter records the deposited energy of particles passing through it. A threshold bounded cluster finder was developed to identify clusters' energy deposition from connected towers of the EEMC(BEMC). The threshold is tuned to discriminate energy deposition of particles produced by collisions from electronic noise and background. Assuming all clusters are photons, neutral pions are identified from cluster pair mass distributions (Figure \ref{pi_mass}). The ratio of the leading tower energy to the cluster energy is selected to be $>$ 0.9 to suppress background such as hadronic showers. To reduce the detector edge bias, there are fiducial volume selections: $1.1<\eta<1.9$ for EEMC and $-0.9<\eta<0.9$ for BEMC. Energy sharing between cluster pairs is defined by the energy difference over the cluster pair energy sum. The energy sharing ratio is chosen to be $<$ 0.7 to reduce non-$\pi^{0}$ background. The mass distributions of cluster pairs are in agreement with previous BEMC $\pi^{0}$ results both in p+p and d+Au runs. Figure \ref{pi_mass} shows the invariant masses of EEMC di-clusters with FMS triggered $\pi^{0}$ in p+p and d+Au data. $\pi^{0}$'s are evident in the EEMC di-clusters.  To emphasize $\pi^{0}$ signals for azimuthal correlations $mass<0.2GeV/c^{2}$ cuts are applied. The FMS $\pi^{0}$ + BEMC $\pi^{0}$ azimuthal correlations with the threshold bounded cluster finder are quantitatively consistent with previous results\cite{for_cen}. To compare with forward+mid rapidity results, we select FMS $\pi^{0}$ $p_{t}^{FMS}>2.5GeV/c$ and EEMC $\pi^{0}$ $1.5GeV/c<p_{t}^{EEMC}<p_{t}^{FMS}$. The azimuthal correlation gives the probability to find a $\pi^{0}$ in the EEMC when a $\pi^{0}$ is found in the FMS. Presently, background and efficiency corrections are not applied. The azimuthal correlations between FMS $\pi^{0}$ and EEMC $\pi^{0}$ are fit with a constant plus a gaussian function. The width of the back-to-back peak is shown in Figure \ref{eemc_pi}. Low statistics in these correlations are caused by cluster pair kinematics. Unlike the BEMC situation,  comparison between data and simulation shows that most of the $\pi^{0}$ candidates are in single clusters in the EEMC. The studies to get more $\pi^{0}$'s using EEMC single clusters with the help of shower maximum detector (SMD) in EEMC\cite{eemcnim} to separate photons is ongoing.
 

\begin{figure}[t]
\centerline{\includegraphics[width=150mm]{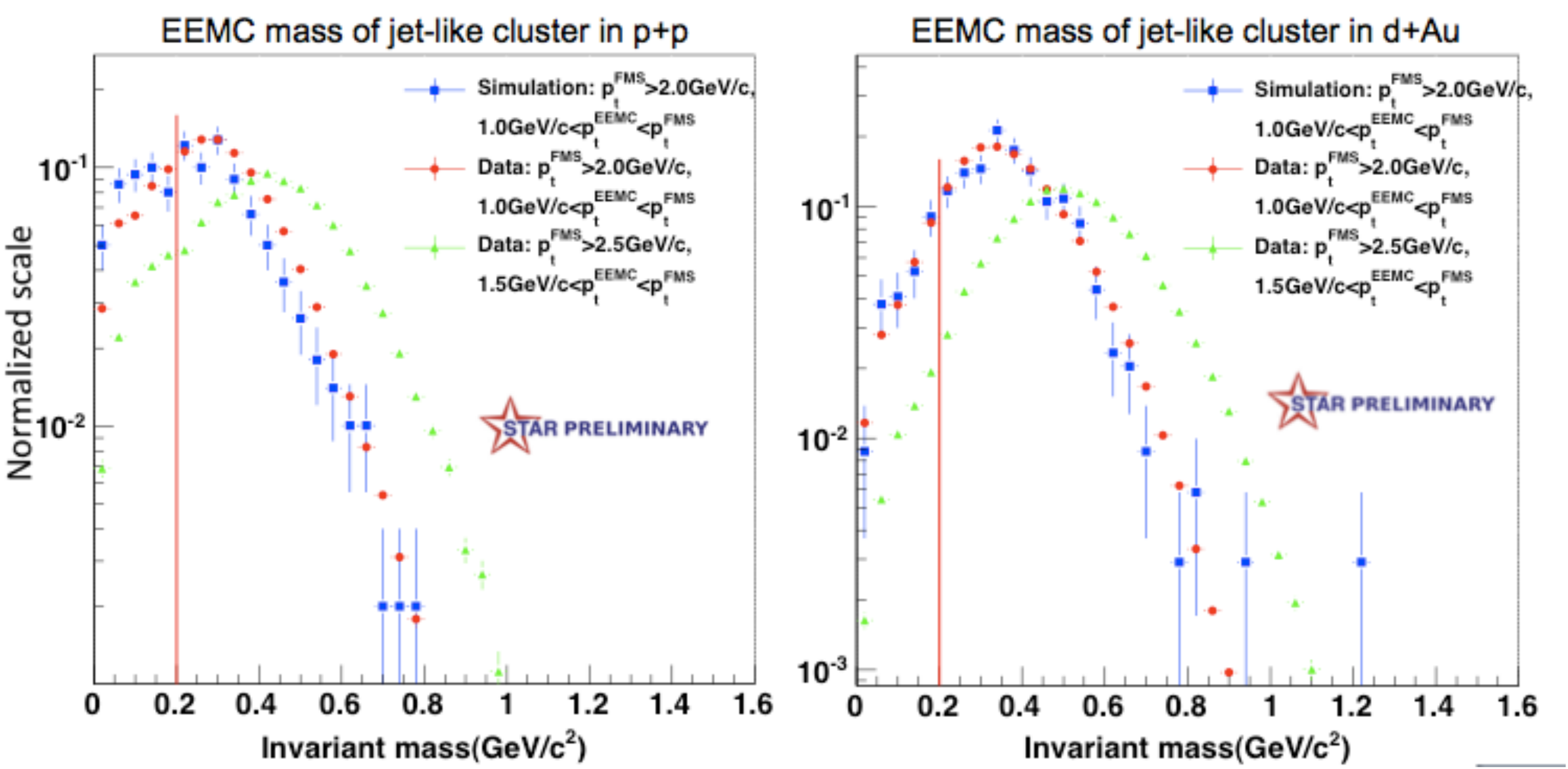}}
\caption{{\it Comparison of data and simulation for EEMC jet-like cluster invariant masses (normalized). With FMS $\pi^{0}$ $p_{t}^{FMS}>2.0GeV/c$ and the EEMC jet-like cluster $1.0GeV/c<p_{t}^{EEMC}<p_{t}^{FMS}$, blue distributions stand for simulation and red distributions stand for data. The green points show the results in data with FMS $\pi^{0}$ $p_{t}^{FMS}>2.5GeV/c$ and the EEMC jet-like cluster $1.5GeV/c<p_{t}^{EEMC}<p_{t}^{FMS}$. The invariant masses of EEMC jet-like clusters in p+p are shown in the left plot and the d+Au results are shown in the right plot. Red lines indicate $0.2GeV/c^{2}$ cut for invariant masses of jet-like clusters. }}
\label{mass}
\end{figure}

\begin{figure}[t]
\centerline{\includegraphics[width=145mm]{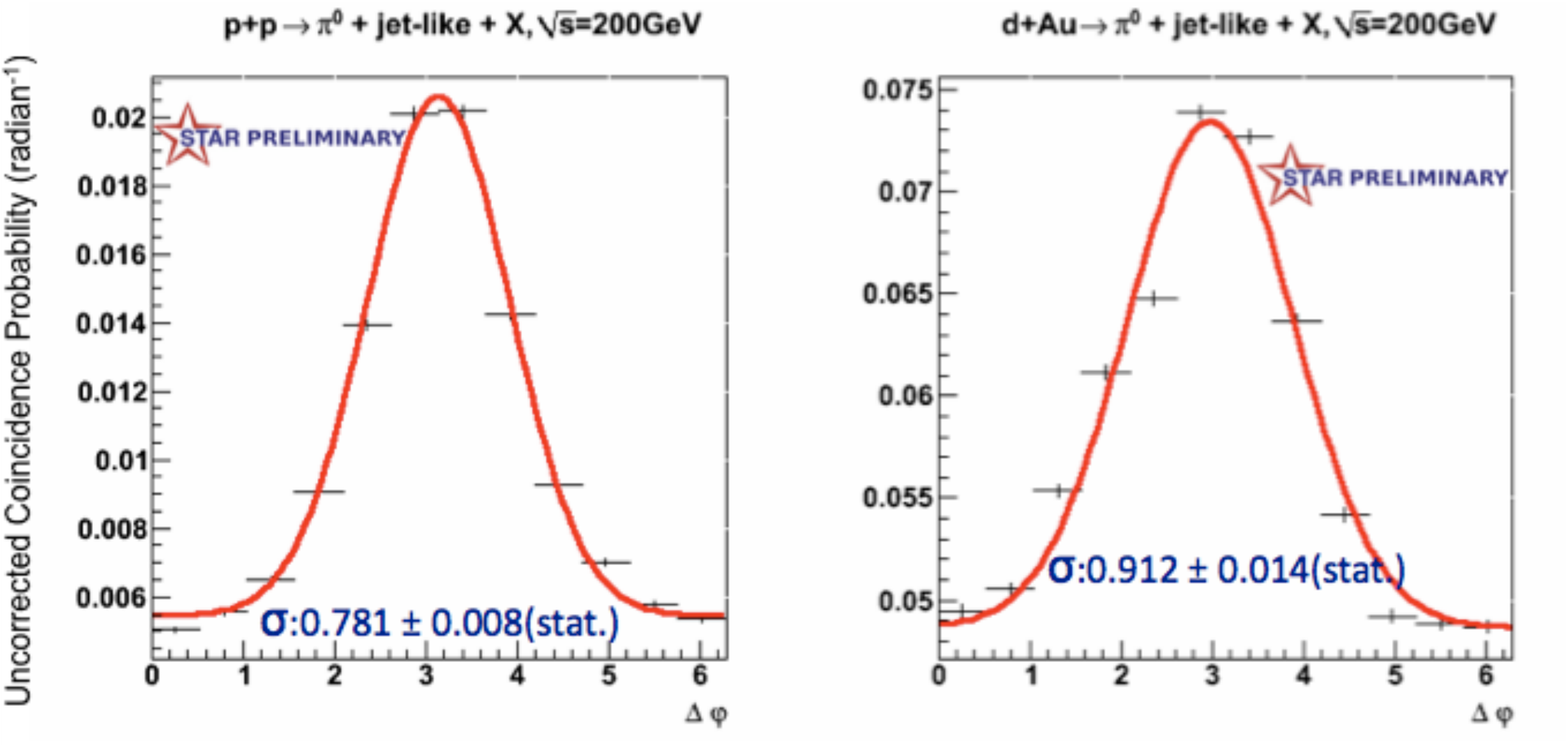}}
\caption{ {\it Uncorrected coincidence probability versus azimuthal angle difference between a forward $\pi^{0}$ ($p_{t}^{FMS}>2.5GeV/c$) and a near-forward rapidity jet-like candidate ($1.5GeV/c<p_{t}^{EEMC}<p_{t}^{FMS}$).  The red lines stand for the fitting functions which are a constant plus a Gaussian function. The left plot shows the correlations in p+p data and d+Au results are shown on the right plot. }}
\label{eemc_jet}
\end{figure}

%% jet-like events
To reduce fluctuations caused by parton fragmentation, jet-like clusters are better substitutes than neutral pions to reflect the behavior of scattered partons. The STAR calorimeters are primarily sensitive to electromagnetic shower energy and only partially sensitive to hadronic shower energy. Consequently, a jet cone algorithm developed from the threshold bounded cluster finder identifies jet-like clusters rather than jets. Initially, we defined a jet-like cluster by using cluster as seed, cone radius $R=0.5$ ($R : \sqrt{\Delta\eta^{2}+\Delta\varphi^{2}}$), and only one iteration. Later on, with more studies including comparisons between data and simulation, we improved the algorithm by using jet cone radius $R=0.6$ and multiple iterations to get stable jet-like clusters.  Triggered neutral pions are detected in the FMS with transverse momentum $p_{t}^{FMS}>2.5GeV/c$ and $2.5<\eta<4.0$. Associated jet-like clusters are detected in the EEMC with transverse momentum $1.5GeV/c<p_{t}^{EEMC}<p_{t}^{FMS}$ and $1.1<\eta<1.9$. Jet-like clusters with $mass>0.2GeV/c^{2}$ are chosen for the correlation analysis to suppress clusters that are only a single $\pi^{0}$, and hence are not explicitly jet-like. The cross section for forward particle productions drops rapidly as $p_{t}$ grows, so lower $p_{t}$ cuts are chosen to compare data and simulation. From the jet-like invariant mass distribution (Figure \ref{mass}), data and simulation are consistent both in p+p and d+Au collisions. The consistency between azimuthal correlations in data and simulation was also checked. The coincident probability (efficiency uncorrected) of azimuthal angle differences between FMS $\pi^{0}$ and EEMC jet-like candidates are shown in Figure \ref{eemc_jet} with FMS $\pi^{0}$ $p_{t}^{FMS}>2.5GeV/c$ and EEMC jet-like candidate $1.5GeV/c<p_{t}^{EEMC}<p_{t}^{FMS}$. The scales of the peak in the back-to-back correlations ($\Delta\varphi=\pi$) are higher in d+Au interaction than that in p+p. Comparisons of azimuthal correlations between p+p and d+Au indicate higher pedestal in d+Au. The width of the peak  centered at $\Delta\varphi=\pi$ in d+Au is significantly broader than that in p+p ($\sigma_{dAu}-\sigma_{pp}= 0.131\pm0.016$). This indicates that the system in d+Au collisions is approaching saturation at the small Bjorken-x region $0.003<x_{BJ}<0.02$. Systematic studies using a random seed in detector coverage and repeating same jet-like clusters algorithm show similar correlation results. Underlying events are also studied with a random seed and a single iteration of the jet finder. These studies show that the jet-like clusters pick up more events in the peak rather than the pedestal compared to the underlying events.

\section{Conclusions and Outlook}
RHIC 2008 run had d+Au collisions at  $\sqrt{s}=200$GeV and the STAR FMS provided opportunities to study forward particle triggered di-hadron (hadron+jet-like) azimuthal correlations to probe the low $x_{BJ}$ region. The FMS-EEMC correlations fill in the region between forward+mid rapidity correlations and forward+forward rapidity correlations to provide a fuller picture of the saturation boundary. Azimuthal correlations between a neutral pion in the FMS and jet-like clusters in the EEMC are found to have a broader back-to-back peak in d+Au collisions than in p+p collisions. Combined with earlier studies, this suggests a smooth transition from low to high parton densities. 

Back-to-back correlation broadening effect from p+p to d+Au is expected to be stronger at lower $p_{t}$ in saturation models. Azimuthal correlations in FMS+EEMC with lower $p_{t}$ cuts are being studied. The forward+forward correlations in d+Au collisions shows dependency on centrality selections\cite{dau_cen}. The STAR Beam-Beam Counter(BBC) multiplicity provides a measure of centrality of the event. Therefore, the azimuthal correlations in d+Au interaction can be divided into peripheral and central part with different multiplicity selections. Comparison between d+Au central collisions and d+Au peripheral collisions will be studied.

%% This is test of bibtex reference.
\section{References}
\bibliographystyle{iopart-num}
\bibliography{xuanli}
\end{document}